%% file: conference.tex
\newcommand{\CLASSINPUTtoptextmargin}{1.9cm}
\documentclass[10pt, conference]{IEEEtran}
\IEEEoverridecommandlockouts

\usepackage{cite}
\usepackage{amsmath,amssymb,amsfonts}
\usepackage{algorithmic}
\usepackage{graphicx}
\usepackage{textcomp}
\usepackage[dvipsnames]{xcolor}
\usepackage{float}
\usepackage{adjustbox}
\usepackage[most]{tcolorbox}
\usepackage{booktabs, multirow, mdframed}
\usepackage{enumitem}
\usepackage{cuted}
\usepackage{url}
\usepackage{circledsteps}
\usepackage{subcaption}
\usepackage[hidelinks,pdfhighlight=/N]{hyperref}
\hypersetup{ %
	pdftitle={Routing Techniques for Error-Corrected Silicon Spin Qubit Quantum Architectures},   %
	pdfsubject={IEEE International Conference on Quantum Computing and Engineering (QCE) 2026}, %
	pdfauthor={Julian Shen, Ludwig Schmid, Robert Wille}   %
}
\newcommand{\ket}[1]{\lvert#1\rangle} %
\def\BibTeX{{\rm B\kern-.05em{\sc i\kern-.025em b}\kern-.08em
    T\kern-.1667em\lower.7ex\hbox{E}\kern-.125emX}}

\usepackage{tikz}

\renewcommand{\tablename}{Tab.}
\renewcommand{\thetable}{\arabic{table}}

\begin{document}
\bstctlcite{IEEEexample:BSTcontrol}

\title{\vspace{-0.4cm}Routing Techniques for Error-Corrected \\ Silicon Spin Qubit Quantum Architectures
}

\author{
    \IEEEauthorblockN{
        Julian Shen\IEEEauthorrefmark{1},
        Ludwig Schmid\IEEEauthorrefmark{1},
        Robert Wille\IEEEauthorrefmark{1}\IEEEauthorrefmark{2}, \vspace{-0.2cm}
    }\\
    \IEEEauthorblockA{
        \IEEEauthorrefmark{1}Chair for Design Automation, Technical University of Munich, Munich, Germany\\
        \IEEEauthorrefmark{2}MQSC, Garching near Munich, Germany\\
        \{julian.shen, ludwig.s.schmid, robert.wille\}@tum.de\\
        \url{https://www.cda.cit.tum.de/research/quantum}
        \vspace{-0.2cm}
    }
}

\maketitle

\begin{abstract}
\input{sections/05_abstract}
\end{abstract}

\input{sections/10_introduction.tex}
\input{sections/20_background}
\input{sections/30_problem_and_idea}
\input{sections/40_implementation}
\input{sections/50_evaluation}

\input{sections/60_conclusion}

\bibliographystyle{IEEEtran}
\bibliography{IEEEabrv,bibliography}

\end{document}

%% file: sections/05_abstract.tex
Silicon spin qubits have emerged as a promising qubit technology due to their favorable scaling and fabrication properties. 
However, efficiently compiling quantum circuits onto spin qubit platforms remains challenging, particularly when accounting for hardware constraints and the high sensitivity to static defects.
Existing compilation approaches for spin qubits either largely ignore error correction, despite its critical role for large-scale quantum computation, or focus on low-level schedule constructions, missing a high-level compilation and routing for logical, error-corrected algorithms.
To address this gap, we introduce a compilation framework for spin qubits based on the recent snakes on a plane model, which utilizes a 2D surface code and qubit teleportation to mitigate errors.
Building on this model, we propose shortest-path and rotation-based algorithms as two novel classes of qubit-routing techniques, along with additional defect-handling and initial-mapping strategies.
We evaluate both algorithms across diverse architectural settings and problem sizes, demonstrating that shortest-path methods excel in sparse, low-defect scenarios, while \mbox{rotation-based} approaches perform better in high-density environments.
An \mbox{open-source} implementation of our framework is publicly available on GitHub as part of the Munich Quantum Toolkit~(MQT) at \url{https://github.com/munich-quantum-toolkit/spin-qubit-routing}.

%% file: sections/10_introduction.tex
\section{Introduction}
\label{sec:introduction}
Quantum computing is steadily progressing towards practical realization. 
As the field continues to advance, research into scalable quantum architectures has explored a diverse range of different qubit technologies~\cite{Kok.2007, Imamoglu.1999, Cirac.1995, Henriet.2020, Kjaergaard.2020}.
Among these, \emph{silicon spin qubits}~\cite{Wild.2012} stand out as a particularly promising candidate due to their favorable manufacturing and scaling properties\mbox{\cite{Zwanenburg.2013, HRL.2026}}, as well as their long coherence times~\cite{Veldhorst.2014} and increasing gate fidelities~\cite{Kawakami.2016, Yoneda.2017, Xue.2022}.

However, in order to utilize these technologies, quantum circuits must first be compiled for the physical hardware before execution. 
This process involves several steps, including initial qubit mapping and routing operations to ensure that all \mbox{two-qubit} interactions are realized in the correct order.
In recent years, compilation approaches for spin qubit platforms have emerged, largely tailored to specific hardware models, such as the crossbar architecture~\cite{Paraskevopoulos.2023, Paraskevopoulos.2024} or shuttling busses~\cite{Escofet.2025, Escofet.2025.3}.
A key limitation of these proposed models is that they focus on compiling physical circuits directly onto the hardware, targeting the NISQ era and neglecting recent developments in \emph{Quantum Error Correction}~(QEC).
Pataki et al.~\cite{patakiCompilingSurfaceCode2025} pioneered QEC compilation for spin qubits, focusing on surface code stabilizer extraction cycles on the crossbar architecture.
Nevertheless, their work still misses a high-level compilation and routing for logical, error-corrected algorithms.
Despite the importance of QEC for scalability, there is currently no compilation framework for spin qubits that explicitly targets routing for error-corrected architectures.

To close this research gap, we explore compilation strategies to optimize qubit routing in spin qubit architectures based on the \emph{snakes on a plane} model recently introduced by Siegel~et~al.~\cite{Siegel.2025}, which provides an intrinsic mechanism for error correction referred to as \emph{snake surgery}.
Based on this framework, we first identify important routing challenges of spin qubits and formalize them using a graph-based abstraction of the original model. 
This formalization transforms the physical routing problem into a classical path-finding problem which can then be tackled with well-known graph algorithms, resulting in two novel routing approaches for spin qubits: \emph{path algorithms} and \emph{rotation algorithms}.
Additionally, we propose two complementary defect-handling mechanisms for robust routing under noise, and we incorporate initial-mapping techniques to reduce the routing overhead.
These components form the proposed framework that compiles quantum circuits into executable schedules for the snakes on a plane model.

In an empirical evaluation, we examine how the proposed algorithms perform across different architectural settings and problem sizes, revealing that path algorithms are better suited for scenarios with sparse qubit density and low-defect architectures, whereas rotation algorithms emerge as the more effective choice in high-qubit-density environments.

In summary, we make the following contributions:
\begin{itemize}
    \item We formalize physical routing for spin qubit architectures as a graph-based path-finding problem, bridging hardware constraints with classical computer science methods.
    \item We propose two new qubit routing strategies: path algorithms and rotation algorithms.
    \item Based on this, we introduce the first high-level routing framework for spin qubit architectures that explicitly incorporates error correction mechanisms.
    \item We empirically evaluate the proposed framework across varying architectural conditions and problem sizes.
\end{itemize}
This paper is structured as follows: \autoref{sec:background} reviews spin qubit quantum computing and the snakes on a plane model.
\autoref{sec:motivation} identifies key routing challenges and formulates the core problem based on a graph formalization.
\autoref{sec:solution} outlines the proposed solution strategies with technical implementation details provided in \autoref{sec:implementation}.
The compilation algorithms are then evaluated in \autoref{sec:evaluation}, and \autoref{sec:conclusion} concludes this paper.

%% file: sections/20_background.tex
\section{Background}
\label{sec:background}
We begin with an overview of spin qubit quantum computers, followed by a discussion of the main error sources in silicon spin qubit devices and a brief review of quantum error correction techniques.
After that, we introduce the snakes on a plane architecture, which serves as the central computational model considered in this work.

\subsection{Spin Qubit Quantum Computers and Compilation}
Spin qubit quantum computers store information in the \emph{spin state} of single electrons confined in a silicon quantum dot~\cite{Murphy.2002}.
The spin can either be in a \mbox{spin-up} or \mbox{spin-down} state, corresponding to the quantum states $\ket{0}$ and $\ket{1}$, respectively. 
Spin qubits are typically realized in \mbox{silicon-germanium} heterostructures or silicon \mbox{metal-oxide-semiconductor} devices, where precise control of the electron’s local environment leads to long coherence times and \mbox{high-fidelity} operations~\cite{Yoneda.2017}. 
Qubit manipulation is achieved using magnetic or electric fields, allowing the realization of both \mbox{single-qubit} and \mbox{two-qubit} gate operations~\cite{Yoneda.2017}. 
Recent experimental results have demonstrated gate fidelities exceeding the thresholds required for quantum error correction in silicon spin qubit platforms~\cite{Xue.2022}.
Measurement in spin qubit systems is typically performed via \mbox{spin-to-charge} conversion~\cite{Loss.1998, Hanson.2007}. In this approach, the quantum dot containing the spin qubit is \mbox{tunnel-coupled} to an electron reservoir. By tuning the gate voltages, only one spin state is energetically allowed to tunnel out of the quantum dot. 
As a result, the spin state is converted into a detectable charge configuration corresponding to the presence or absence of an electron. 
This technique enables \mbox{high-fidelity} readout and is well suited for scalable \mbox{silicon-based} quantum dot architectures~\cite{Hanson.2007}.

However, performing \mbox{two-qubit} gates in spin qubit architectures requires the participating qubits to be physically close to one another, making qubit shuttling necessary.
One promising shuttling approach is the conveyor belt technique, which creates a moving confinement potential that can trap and continuously transport single electrons across the device~\cite{Seidler.2022}.
Beyond the physical realization of qubit transport, several abstract models for spin qubit quantum computers have been proposed~\cite{Crawford.2022, Paraskevopoulos.2023, Paraskevopoulos.2024, Huang.2025, Escofet.2025}. 
These models enable the development of compilation techniques that compute efficient schedules for qubit movements, ensuring that required qubit interactions occur in the correct order while minimizing routing overhead.
However, none of the referenced models address errors or provide mechanisms for detecting or correcting them, leaving the compilation methods equally unaware.

\subsection{Errors in Spin Qubit Quantum Computers}
\label{sec:errors}
The most detrimental source of errors to be considered in silicon spin qubit devices are \emph{charge noises}~\mbox{\cite{Culcer.2009, Yoneda.2017, Shehata.2023}}.
Charge noise arises from random fluctuations of charges between defects spread across the device and effectively induces uncontrolled variations in the magnetic field landscape, leading to dephasing errors.
In general, charge noise can be divided into the following two categories:
\begin{enumerate}
    \item \textbf{Abrupt Charge Disturbance}: Local charge rearrangements that cause sudden and severe phase errors.
    \item \textbf{Small-Amplitude Fluctuation}: Slow and gradual variation in the magnetic field landscape that leads to mild but persistent dephasing.
\end{enumerate}
One common approach to mitigate such errors is the use of \emph{Quantum Error-Correcting Codes} (QECCs). 
QECCs protect quantum information by encoding the qubits of a quantum circuit, referred to as \emph{logical qubits}, into entangled states of multiple \emph{physical qubits}. 
This redundancy allows certain errors affecting individual physical qubits to be detected and corrected without directly measuring or disturbing the encoded logical quantum state.
Many QECCs are formulated within the \emph{stabilizer formalism}~\cite{Gottesman.1997}, where a set of commuting operators, called stabilizers, define the valid subspace of the encoded logical qubit. 
By repeatedly measuring these stabilizers, one obtains an error syndrome that reveals the presence and location of errors~\cite{Kolmogorov.2009} while preserving the logical information.
A prominent example of such a code is the \emph{surface code}~\cite{Fowler.2012}, which arranges qubits on a \mbox{two-dimensional} lattice and uses local stabilizer measurements to detect both \mbox{bit-flip} and \mbox{phase-flip} errors. 
Surface codes are particularly attractive for \mbox{near-term} quantum hardware, including spin qubit platforms, because they only require \mbox{nearest-neighbor} interactions and exhibit relatively high error thresholds.

Recently, Siegel et al.~\cite{Siegel.2025} have introduced a new architectural model for spin qubit quantum computers called \emph{snakes on a plane}. 
This model explicitly incorporates \mbox{surface-code-based} error correction to mitigate the aforementioned noise sources and employs a dedicated protocol for \mbox{fault-tolerant} qubit transport, called \emph{snake surgery}, as discussed in the following.

\subsection{Snakes on a Plane Model}
\label{sec:snakes-on-a-plane}
\begin{figure}[tb]
    \centering
    \subfloat[General layout.\label{fig:model-overview}]{
        \centering
        \includegraphics[width=0.45\linewidth]{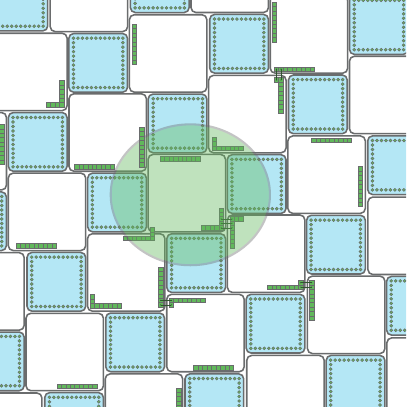}}
    \hfill
    \subfloat[Close-up of the green region.\label{fig:close-up}]{
        \centering
        \includegraphics[width=0.45\linewidth]{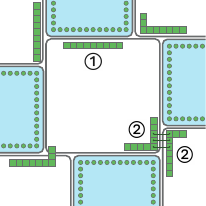}
    }
    \caption{Sketch of the snakes on a plane architecture~\cite{Siegel.2025}.}
    \label{fig:general_layout}
\end{figure}

In the snakes on a plane model~\cite{Siegel.2025}, the device architecture is modeled as a 2D latticework formed of \mbox{low-dimensional} array strands that occasionally meet at junctions.
A schematic overview of the architecture is provided in \autoref{fig:general_layout}.
Error correction based on surface codes is implemented using static ancilla qubits (green dots) located in blue squares, which facilitate the required stabilizer measurements.
Logical qubits are built from multiple physical qubits (green squares), creating snakes that can be shuttled along the inner edges of the white squares in the latticework.
When a snake reaches a junction, it can either remain in its current square or move to a neighboring square.
In general, snakes move for two main reasons in this model:
\begin{itemize}
    \item \emph{Error Correction}: When a snake moves along the outer edge of a blue square, referred to as a \emph{stabilizer edge}, its physical qubits face the ancilla qubits located along the edge.
    Stabilizer measurements can then be performed between pairs of qubits that are opposite to each other~\cite{Siegel.2025}.
    By repeatedly shuttling the snake back and forth along a stabilizer edge and performing the stabilizer measurements in an appropriate sequence, a surface code can be implemented and the logical qubit represented by the snake remains \mbox{error-free}.
    \begin{description}[topsep=6pt, leftmargin=0pt]
        \item[Example 1.] \itshape The snake denoted with \raisebox{.5pt}{\textcircled{\raisebox{-.8pt} {\small{1}}}} in \autoref{fig:close-up} is being shuttled along a stabilizer edge while performing stabilizer measurements using the green ancilla qubits.
    \end{description}
    \item \emph{Interaction}: Two snakes can be moved to an edge where they face each other, enabling them to interact and realize a logical (transversal) \mbox{two-qubit} gate.
    Edges, where snakes can be brought into parallel alignment, are therefore referred to as \emph{interaction edges}.
    \begin{description}[topsep=6pt, leftmargin=0pt]
        \item[Example 2.] \itshape  The two snakes denoted with \raisebox{.5pt}{\textcircled{\raisebox{-.8pt} {\small{2}}}} in \autoref{fig:close-up} interact with each other at an interaction edge. Qubit interactions (black connections) can be performed whenever the respective qubits are opposite to each other.
    \end{description}
\end{itemize}
If a snake permanently remains at a stabilizer edge, errors can be mitigated through repeated stabilizer measurements.
However, when a snake travels over a larger distance, additional mechanisms are required to handle the two noise types described previously in \autoref{sec:errors}.

\textbf{Abrupt charge disturbances} are modeled as defective edges that can occur within the snakes on a plane architecture. 
Each functioning edge may become defective with probability $p_\text{defect}$, while each defective edge recovers with probability $p_\text{recovery}$, which models the natural decay of dephasing effects on shuttling links.
If a snake traverses a defective edge, it acquires an error and, therefore, does not arrive at its destination in an \mbox{error-free} state.
In this case, the shuttling operation can be reversed using a dedicated \mbox{defect-handling} protocol called \emph{snake surgery}~\cite{Siegel.2025}.
In this protocol, the length of the snake is first doubled before shuttling using lattice surgery~\cite{Horsman.2012}, creating an entangled pair consisting of two parts:
\begin{itemize}
    \item \emph{Tail}: The original snake, which remains at the initial stabilizer edge and continues to be stabilized and protected against errors.
    \item \emph{Head}: A newly created extension of the snake that is shuttled across the device towards the destination.
\end{itemize}
Once the head reaches its destination, a measurement of either the head or the tail determines where the logical state is teleported to. If the route was \mbox{error-free}, the state is teleported to the head, otherwise, it is teleported back to the tail, effectively reversing the shuttling process.
\pagebreak

\begin{figure}[tb]
    \centering
    \includegraphics[width=0.5\linewidth]{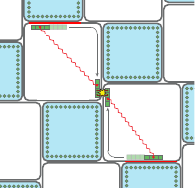}
    \caption{Illustration of the snake surgery protocol~\cite{Siegel.2025}.}
    \label{fig:snake-surgery}
\end{figure}

\begin{description}[topsep=6pt, leftmargin=0pt]
    \item[Example 3.] \itshape Consider the scenario illustrated in \autoref{fig:snake-surgery}. To enable the interaction between the two snakes, snake surgery is performed on each snake, splitting it into an entangled head-tail pair, indicated by squiggly lines. The heads are then shuttled to the designated interaction edge, while the tails remain stabilized at their original positions.
\end{description}
Since snake surgery requires the detection of errors in order to determine when the shuttling process must be reversed, the snakes on a plane model further incorporates two dedicated mechanisms for detecting defects~\cite{Siegel.2025}:
\begin{itemize}
    \item \emph{Monitor Qubits}: These are additional qubits which are prepared in the $\ket{+}$ state and then interlaced with the data qubits of a snake.
    The qubits are then shuttled together, ideally leaving the monitor qubits unchanged, but defects on the device may cause them to accumulate a phase during the process.
    In this case, measuring these monitor qubits in the $X$-basis allows us to detect whether the snake has encountered a defect or not.
    \item \emph{Complementary Gap}: This leverages the fact that ancilla qubits continuously gather information about potential errors by measuring stabilizers throughout the shuttling process. 
    The complementary gap~\cite{Gidney.2023} characterizes the confidence of a surface code decoder that it can determine the cause of an observed error.
    In case of a high-risk event, snake surgery can be used to repeat the shuttling process and to ensure logical correctness.
\end{itemize}
\textbf{Small-amplitude fluctuations} can be mitigated using the \emph{singlet-triplet qubit encoding}~\cite{Levy.2002}, where a logical qubit is encoded into two electron spins. 
Hereby, the logical states depend on the relative spin orientation rather than the absolute spin directions of the electrons, making \mbox{singlet-triplet} qubits far less sensitive to uniform magnetic-field fluctuations~\cite{Loss.1998}.

In this contribution, we adopt the snakes on a plane model to represent an error-correcting architecture and focus on the routing problem arising from snake surgery.

%% file: sections/30_problem_and_idea.tex
\section{The Snake Routing Problem}
\label{sec:motivation}
Following the snakes on a plane model reviewed above, which incorporates error correction as an intrinsic feature, the problem of efficient compilation still remains open.
In this work, we focus on two-qubit gates that require the participating snakes to be physically close to each other.
Other \mbox{QEC-related} compilation tasks, e.g., magic-state creation and distribution~\cite{Bravyi.2005}, are not the focus of this work but are interesting follow-up questions that can be addressed using the results presented here.
Consequently, the compilation in this setting reduces to a routing problem.
This section begins by outlining physical routing challenges and then introduces a graph-based formalization, enabling the problem to be formulated as a \mbox{path-finding} problem.
This formalization allows the routing task to be tackled using established algorithms.

\subsection{Routing Challenges}
\label{sec:challenges}
\subsubsection{Snake Collision}
Since snakes are not permitted to overtake one another, they may collide during shuttling operations. Preventing such collisions requires careful coordination of their movements, particularly in densely populated grids with a large number of snakes.

\begin{figure}[tb]
    \centering
    \subfloat[Snake collision.]{
        \includegraphics[width=0.35\linewidth]{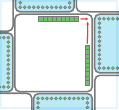}
        \label{fig:snake-collision}
    }
    \hfill
    \subfloat[Correct handling of defective edges.]{
        \includegraphics[width=0.55 \linewidth]{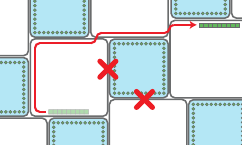}
        \label{fig:snake-rerouting}
    }
    \caption{Routing challenges in the snakes on a plane model~\cite{Siegel.2025}.}
    \label{fig:routing-conflicts}
\end{figure}

\begin{description}[topsep=6pt, leftmargin=0pt]
    \item[Example 4.] \itshape Consider the scenario illustrated in \autoref{fig:snake-collision}. If both snakes move simultaneously in the indicated directions, they will collide with each other. To avoid this conflict, one of the snakes must wait until the other one has cleared the intended path before proceeding.
\end{description}

\subsubsection{Tail Blockage of Stabilizer Edges}
When a snake is shuttled over a larger distance, its tail remains anchored at the originating stabilizer edge. 
This allows continuous stabilizer measurements to be performed on the tail, ensuring that the shuttling process can be repeated if the head encounters any error.
Consequently, the originating stabilizer edge remains occupied and cannot be used to route other snakes until the tail is teleported away.

\begin{description}[topsep=6pt, leftmargin=0pt]
    \item[Example 5.] \itshape Consider the scenario illustrated previously in \autoref{fig:snake-surgery}. As a result of the snake surgery protocol, the red edges remain occupied by the stationary tails and are therefore blocked for routing other snakes until the tails are teleported away.
\end{description}

\subsubsection{Defect Handling}
Edges that are known to be defect should be avoided during shuttling in order to reduce the need for snake surgery teleportations and to minimize the required routing overhead.

\begin{description}[topsep=6pt, leftmargin=0pt]
    \item[Example 6.] \itshape Consider the scenario illustrated in \autoref{fig:snake-rerouting}. The red crosses indicate defective shuttling links that must be avoided. This can be achieved by routing the snake along the red path.
\end{description}

\pagebreak
\subsection{Graph Formalization and Problem Formulation}
\label{sec:graph-and-problem}
\begin{figure}[tb]
    \centering
    \vspace{-1mm}
    \includegraphics[width=0.7\linewidth]{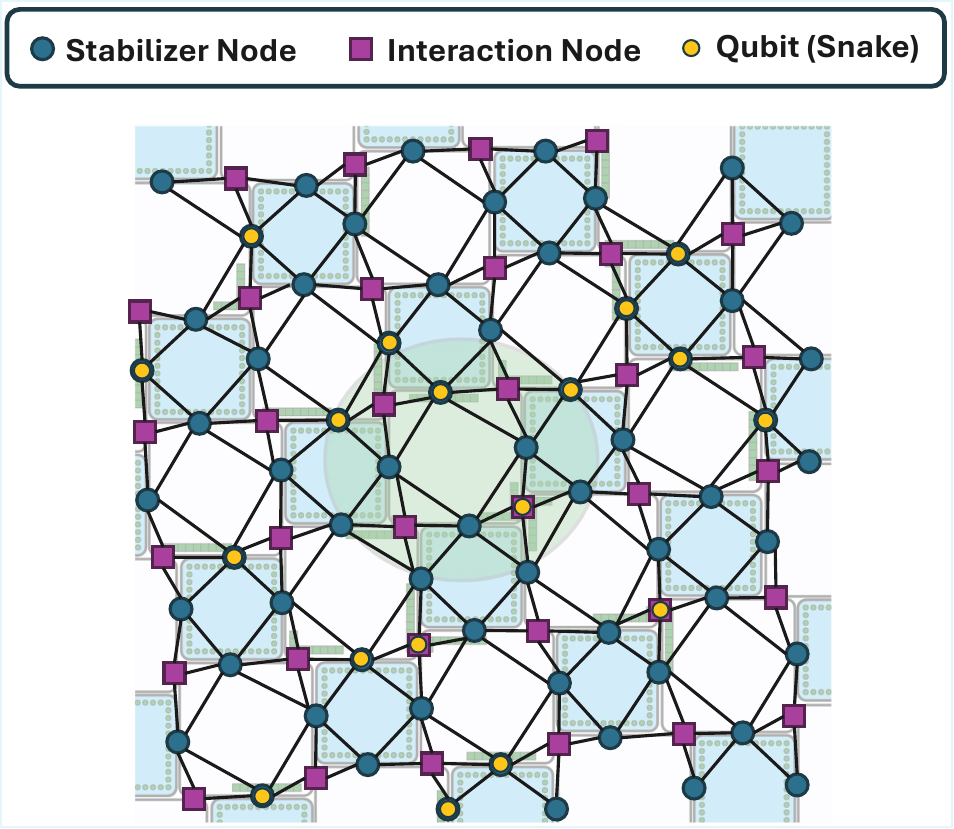}
    \caption{The graph formalization process starting from the snakes on a plane model as shown in \autoref{fig:model-overview}.}
    \label{fig:graph-creation}
\end{figure}

To systematically address the snake routing problem, we formalize the underlying architecture into a graph representation, allowing the physical routing problem to be formulated as a path-finding problem.
This formalization process is illustrated in \autoref{fig:graph-creation}.
Based on the physical architecture shown in \autoref{fig:general_layout}, the main elements of the snakes on a plane model are mapped to graph elements as follows:
\begin{itemize}
    \item Snakes are represented as single qubits (yellow circles) that can occupy exactly one node at any given time.
    \item Stabilizer edges become blue circular nodes, called \emph{stabilizer nodes}. Stabilizer nodes can be occupied by at most one snake at a time. Occupying a stabilizer node corresponds to performing stabilizer measurements.
    \item Interaction edges become purple square nodes, called \emph{interaction nodes}. These nodes are only occupied when two snakes intend to perform an interaction. 
    Accordingly, interaction nodes have a capacity of two and must be occupied by exactly two snakes simultaneously to realize a two-qubit interaction.
\end{itemize}
The edges of the graph are defined in accordance with the snakes on a plane model and each edge can only be traversed by at most one qubit at each point in time.
In particular, edges between stabilizer nodes represent the option to shuttle a snake along stabilizer edges within a square of the original architecture without engaging in an interaction. 

Similar to the original snakes on a plane model, defects on the device are modeled by defective edges that can occur with a certain probability $p_\text{defect}$ during the shuttling process and recover with probability $p_\text{recovery}$.
If a qubit traverses a defective edge along its route, it collects an error and is teleported back to its original starting position via snake surgery upon reaching its destination.
Therefore, the starting positions of shuttled qubits may only be occupied by other qubits if the shuttled qubit undergoes a snake surgery teleportation to its destination. 
Otherwise, these starting nodes remain occupied by the respective tails.

Based on this formalization and the physical challenges reviewed in \autoref{sec:challenges}, we derive the following routing problem:
\begin{tcolorbox}[title=Snake Routing Problem, colback=white, colframe=black, breakable]
Given an initial placement of qubits on stabilizer nodes and a sequence of required qubit interactions, determine a routing schedule that realizes these interactions while optimizing routing performance and satisfying the following properties:
\begin{enumerate}
    \item The capacities of nodes and edges are respected.
    \item Qubits do not collide with other qubits or their respective tails.
    \item Qubits remain on stabilizer nodes whenever possible to enable error correction through stabilizer measurements. They may only leave stabilizer nodes for the time span of their qubit interaction.
    \item Only qubits that are error-free may interact. Defective qubits are routed again using snake surgery. Noisy edges have to be avoided.
    \item The logical order of interactions is respected.
\end{enumerate}
\label{misc:problem-statement}
\end{tcolorbox}

\section{Proposed Solution}
\label{sec:solution}
Having established a graph-based formalization of the snake routing problem, it  can now be tackled using \mbox{well-known} graph algorithms. 
We propose two classes of algorithms suitable for this problem: \emph{path algorithms} and \emph{rotation algorithms}.
Their robustness and efficiency are further improved through the use of defect-handling mechanisms and initial-mapping techniques.
Together, these components form the compilation framework proposed in this work, enabling the automatic translation of quantum circuit instructions into executable schedules for the snakes on a plane architecture.
As discussed above, we assume that magic-state preparation and decomposition are already resolved and focus on the routing task for the corresponding logical qubits.
As a result, the input to our compilation flow is a sequence of interactions between given logical qubits.

\subsection{Path and Rotation Algorithms}
\label{sec:path-and-rotation}
\begin{figure}[tb]
    \centering
    \subfloat[Example for local circumvention used by path algorithms.\label{fig:rerouting-scheme}]{
        \includegraphics[width=0.40\linewidth,angle=0.7]{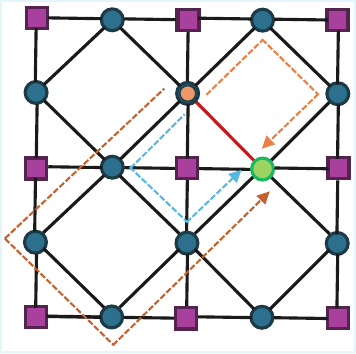}

    }
    \hfill
    \subfloat[Example for dynamic cycles used by rotation algorithms.\label{fig:dynamic-cycles-scheme}]{
        \includegraphics[width=0.45\linewidth,angle=0.7]{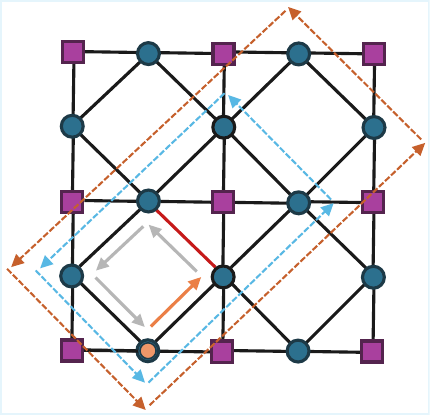}
        
    }
    \caption{Illustration of dynamic path adaption schemes.}
\end{figure}

This section describes the workflow of the proposed path and rotation algorithms used to perform qubit routing on the snakes on a plane model.
Both algorithms share a common structure, consisting of five major building blocks that transform a given sequence of qubit interactions into an executable shuttling schedule. 
The main differences between the two approaches lie in how they resolve path obstructions and handle defects.
The general idea of both algorithms can be summarized into the following steps:
\begin{enumerate}
    \item \textbf{Layer Creation}: To enable the simultaneous shuttling of qubits without violating the prescribed interaction logic, the algorithms first partition the given sequence of interactions into layers. 
    \item \textbf{Path Computation}: For each interaction in a layer, a Breadth-First Search (BFS) is executed to identify the interaction node that minimizes the summed distance to both qubits. This node is called the \emph{meeting node} of that qubit pair. Furthermore, qubits are classified either as \emph{active qubits} or \emph{idle qubits}, depending on whether they participate in an interaction in the current layer or not. During this step, idle qubits are ignored in order to compute the most efficient paths for active qubits to their meeting nodes using an $A^*$ search.
    \item \textbf{Relocation of Idle Qubits}: Since idle qubits were not considered during path computation in the previous step, some of the paths may be obstructed by idle qubits. 
    To resolve this, these \emph{blocking idle qubits} have to be relocated to clear the paths. This is where the two algorithms differ from each other:
    \begin{itemize}
        \item Path algorithms first relocate blocking idle qubits to the nearest free stabilizer nodes that are not on any other paths, before any active qubits are moved.

        \begin{description}[topsep=6pt, leftmargin=0pt]
            \item[Example 7.] \itshape Consider the scenario illustrated in \autoref{fig:path-scheme}. Idle qubits are shown in gray, while the two pairs of active qubits and their respective paths to the meeting nodes are highlighted in green and orange. Idle qubits located along these paths are relocated to the nearest available stabilizer nodes.
        \end{description}

        \item Rotation algorithms relocate idle qubits simultaneously by rotating them along diamond-shaped cycles.

        \begin{description}[topsep=6pt, leftmargin=0pt]
            \item[Example 8.] \itshape Consider the scenario illustrated in \autoref{fig:rotation-scheme}. The orange qubits are scheduled to interact at the central interaction node and must first move onto the green nodes to reach it. Idle qubits blocking these paths are rotated away along diamonds, simultaneously to the movements of the orange qubits.
        \end{description}
       This simultaneous relocation is only possible if snake surgery teleportation is performed at every hop.
       Without teleporting the tail at each step, previously occupied nodes would remain blocked, preventing other qubits from moving onto them. 
       Continuously teleporting the tail removes tail blockage entirely, at the cost of an increased number of teleportation operations.
    \end{itemize}
    \begin{figure}[tb]
        \centering
        \vspace{-1.5mm}
        \subfloat[Path algorithm scheme.]{
            \includegraphics[width=0.48\linewidth, ]{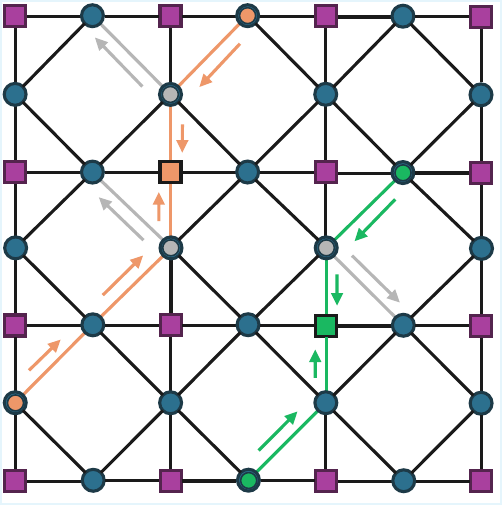}
            \label{fig:path-scheme}
        }
        \hfill
        \subfloat[Rotation algorithm scheme.]{
            \includegraphics[width=0.38\linewidth,angle=0.65]{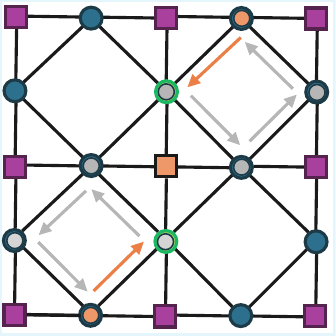}
            \label{fig:rotation-scheme}
        }
        \caption{Illustration of path and rotation schemes.\vspace{-2mm}}
    \end{figure}
    \item \textbf{Defect Handling}: Before the qubits are physically shuttled, the defect model samples all edges. Edges that become defect along the paths of active or idle qubits are handled by the following defect-handling strategies:
    \begin{itemize}
        \item \textbf{Waiting}: The blocked qubit remains at its current node and waits until the defective edge has recovered. This strategy can be used by both algorithms, referred to as the \emph{path} and \emph{rotation algorithm with waiting}.
        \item \textbf{Dynamic Adaption}: Path algorithms can try to reroute blocked qubits using a local circumvention, referred to as the \emph{path algorithm with rerouting}. 
        \begin{description}[topsep=6pt, leftmargin=0pt]
            \item[Example 9.] \itshape In \autoref{fig:rerouting-scheme}, the orange qubit tries to reach the green node but is blocked by the red defective edge. Path algorithms can try to reroute the qubit around the defect along the illustrated circumvention paths.
        \end{description}
        Rotation algorithms can try to avoid defective links by shuttling qubits along larger cycles rather than the default diamond-shaped paths, referred to as the \emph{rotation algorithm with dynamic cycles}.
        \begin{description}[topsep=6pt, leftmargin=0pt]
            \item[Example 10.] \itshape In \autoref{fig:dynamic-cycles-scheme}, the orange qubit is scheduled to move along the highlighted cycle, which contains a defective edge (red). Rotation algorithms may instead route the qubit along a larger cycle, e.g. blue or brown, to bypass this faulty edge.
        \end{description}
        However, if these dynamic path adoptions also fail, the waiting mechanism has to be used as a fallback.
    \end{itemize}
    \item \textbf{Layer Execution}: In the last step, the shuttling plan is executed for the current layer. The procedure then continues with the next layer, restarting from Step 2.
\end{enumerate}
Once a complete routing schedule has been determined on the graph formalization, the corresponding qubit movements can be directly mapped back to snake movements in the snakes on a plane model.
Hence, this translation yields a complete routing plan for the physical architecture.

\subsection{Initial Qubit Mapping}
It has been shown that the initial qubit mapping can have a substantial impact on the efficiency of the final routing schedule~\cite{Siraichi.2018, Zulehner.2018}.
Consequently, the proposed framework offers the following three \mbox{initial-mapping} techniques:
\begin{itemize}
    \item \textbf{Random Mapping (Baseline)}: The logical qubits from the quantum circuit are randomly assigned to physical qubits on the device. %
    \item \textbf{Reverse Traversal Mapping}~\cite{Li.2019}: Start with a random mapping and execute the reversed quantum circuit on that mapping. The resulting qubit placement of the execution is then used as the initial mapping for the original quantum circuit.
    \item \textbf{Interaction-based Mapping}~\cite{Park.2022}: First, the qubit with the highest number of interactions is placed at the center of the graph. Qubits that interact with it are then placed around it recursively, ensuring that frequently interacting qubits are mapped to nearby locations.
\end{itemize}

\subsection{Resulting Compilation Flow}
\begin{figure}[t]
    \centering
    \includegraphics[width=\linewidth]{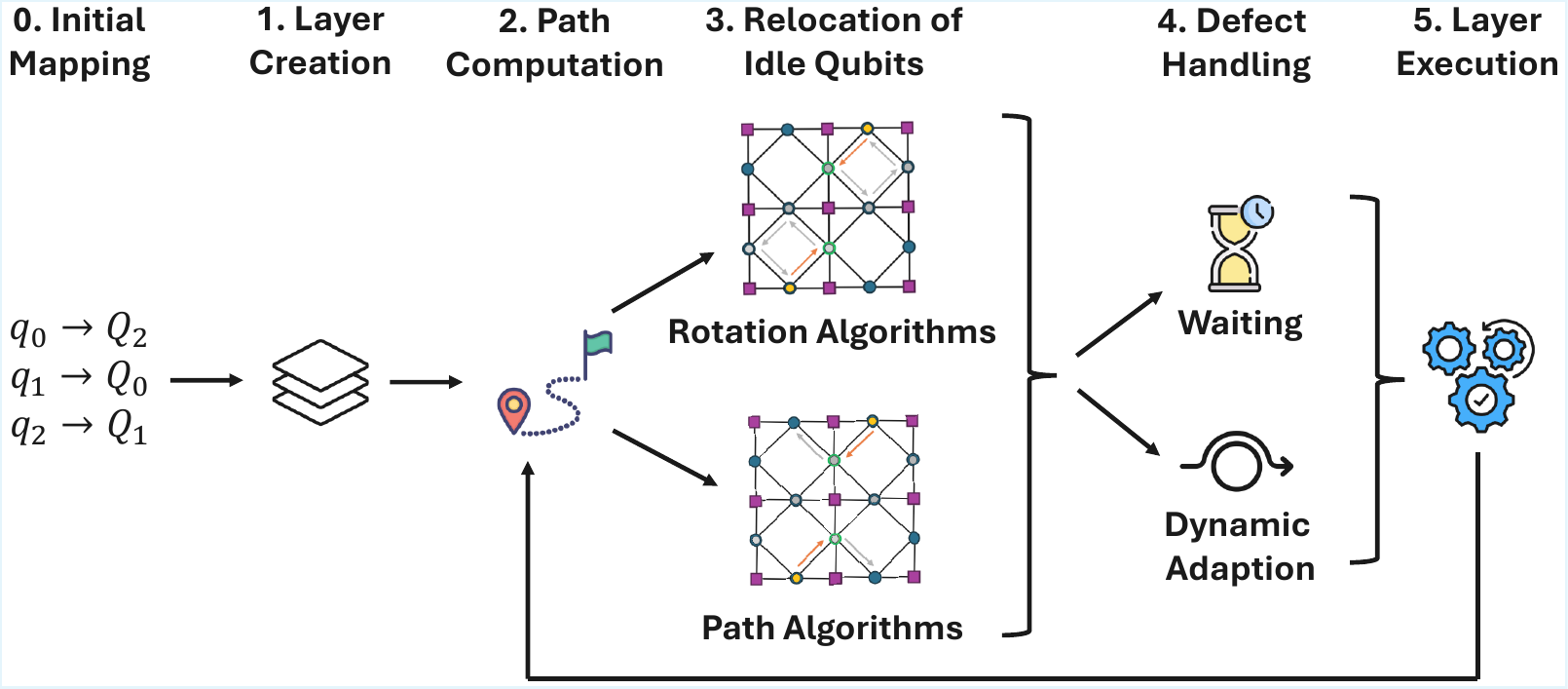}
    \caption{Overview of the complete compilation flow.}
    \label{fig:overview}
\end{figure}

By combining the presented initial-mapping techniques (as Step 0) with the proposed routing algorithms, we obtain a complete compilation framework for the snakes on a plane architecture. 
This framework automatically translates quantum circuits into executable shuttling schedules for spin qubit devices while respecting the physical constraints of the hardware.
The complete compilation flow of the proposed framework is shown in \autoref{fig:overview}.

%% file: sections/40_implementation.tex
\section{Implementation Details}
\label{sec:implementation}
After presenting the core structure and fundamental ideas of the proposed compilation algorithms, we now examine in greater detail how  the key concepts are implemented at a technical level.
In particular, computing collision-free paths for active qubits to their meeting nodes and determining relocation paths for idle qubits require careful algorithmic design to ensure efficient and collision-free routing under the architectural constraints of the snakes on a plane model.

\subsection{Path Constraints}
\label{sec:constraints}
A primary objective of the routing algorithms is to prevent collisions between qubits during shuttling operations. Furthermore, highly congested configurations should be avoided, as they increase the complexity of qubit relocation and reduce the efficiency of the routing process.
To address these challenges, we introduce the following path constraints that limit potential collision scenarios and reduce congestion during routing:
\begin{enumerate}
    \item \emph{Unique Meeting Node}:
    Each active qubit pair within a layer is assigned a distinct meeting node. This prevents congestion and potential collisions in the vicinity of interaction nodes.
    \item \emph{Synchronization of Interactions}: Since the distances from the two qubits of a pair to their meeting node may differ, their arrival times may be misaligned. To synchronize the interaction, the qubit that arrives earlier waits at the stabilizer node immediately preceding the meeting node. We define this waiting point as its \emph{target node}.
    \begin{figure}[tb]
        \centering
        \includegraphics[angle=-90, width=0.5\linewidth]{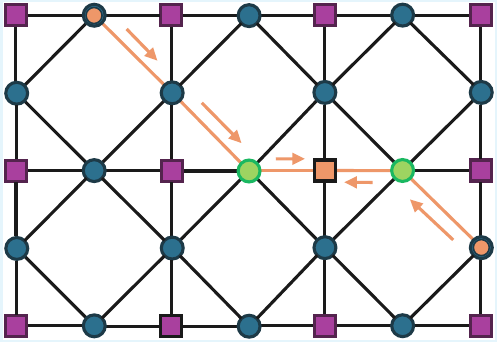}
        \caption{Example for interaction synchronization.}
        \label{fig:path-sync}
    \end{figure}

    \begin{description}[topsep=6pt, leftmargin=0pt]
        \item[Example 11.] \itshape Consider the scenario illustrated in \autoref{fig:path-sync}. The path of the right orange qubit to the meeting node is shorter than the path of the left orange qubit. Consequently, the right qubit must wait at its target node (right green node) until the left qubit reaches its corresponding target node (left green node). Once both qubits have arrived, they can proceed to the interaction node to perform the interaction.
    \end{description}
    \item \emph{Unique Target Node}:
    Each active qubit is assigned a distinct target node to reduce congestion.
    Additionally, qubits are moved back to their target nodes after interaction. 
    If two qubits had the same target node, this would lead to a collision.
\end{enumerate}
The above constraints must be satisfied by both path and rotation algorithms. However, due to the differences in qubit relocation, the last constraint is \mbox{algorithm-specific}.
Path algorithms teleport tails only when the corresponding heads reach their targets, in order to reduce the number of teleportations. 
Consequently, the following constraint must be satisfied:
\begin{enumerate}
\setcounter{enumi}{3}
\item (Path Alg.) \emph{Tail Blockage}:
A routing path must not pass through the initial positions of other qubits that are being routed simultaneously, as these nodes remain occupied by stationary tails during the shuttling process.
\end{enumerate}
In contrast, rotation algorithms employ a cyclic relocation scheme which requires \mbox{snake surgery} teleportation to be performed continuously to free occupied nodes. 
As a result, tail blockage is eliminated entirely.
However, this introduces a different constraint:
\begin{enumerate}
\setcounter{enumi}{3}
\item (Rotation Alg.) \emph{Rotation Conflicts}:
Qubits routed simultaneously must not be assigned to the same rotation cycle, as this would lead to conflicts. 
Furthermore, neighboring cycles must also be rotated sequentially.

\begin{description}[topsep=6pt, leftmargin=0pt]
    \item[Example 12.] \itshape Consider the scenario illustrated in \autoref{fig:single-cycle-conflict}. The orange and green active qubits rotate simultaneously within the same cycle, leaving no opportunity for the gray qubits to relocate. In such cases, the active qubits must be shuttled sequentially.
\end{description}

\begin{description}[topsep=6pt, leftmargin=0pt]
    \item[Example 13.] \itshape Consider the scenario illustrated in \autoref{fig:two-cycle-conflict}. The orange and green active qubits rotate simultaneously in neighboring cycles, forcing the central gray idle qubit to relocate in two conflicting directions at once. This conflict is resolved by executing the rotations sequentially rather than in parallel.
\end{description}
\begin{figure}[tb]
    \centering
    \vspace{-0.7mm}
    \subfloat[Single cycle conflict.]{
        \makebox[0.3\linewidth][c]{%
            \resizebox{0.23\linewidth}{!}{\includegraphics{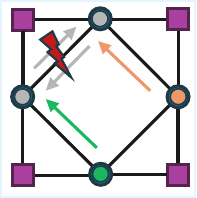}}%
        }
        \label{fig:single-cycle-conflict}
    }
    \hspace{0.5em}
    \subfloat[Neighboring cycles conflict.]{
        \makebox[0.53\linewidth][c]{%
            \resizebox{0.4\linewidth}{!}{\includegraphics{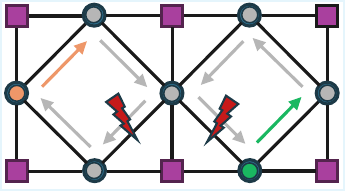}}%
        }
        \label{fig:two-cycle-conflict}
    }
    \caption{Examples for rotation conflicts.}
\end{figure}

\end{enumerate}
The path computation procedure described in the following section is implemented in a way that it adheres to all of these path constraints.

\subsection{Collision-Free Path Computation}
\label{sec:path-comp}
To compute collision-free paths to the meeting nodes in Step~2 of the workflow illustrated in \autoref{sec:path-and-rotation}, both algorithms utilize an adapted $A^*$ search in combination with a reservation table that maintains the current routing state.
Each qubit is represented by a state $(v, t)$, indicating that the qubit occupies node $v$ at timestep $t$.
The reservation table records which nodes and edges are occupied at each timestep. 
It is updated whenever a qubit changes its state, ensuring that conflicts between qubits are avoided during path planning.
The cost of a path is defined as the tuple $(moves,time)$, capturing both the number of qubit moves and the elapsed time.
Since physical qubit movement is more expensive than waiting, the number of moves is prioritized over time when comparing costs.
To fully specify the adapted $A^*$ search, we define the following components, in accordance with the standard formulation from Ref.~\cite{Hart.1968}:
\begin{itemize}
    \item \emph{Successor Function}: From a qubit state $(v,t)$, a qubit can either wait or move to a neighboring node. This leads to the following successor transitions that are possible:
    \begin{itemize}
        \item Wait: $(v,t)\!\rightarrow\!(v,t\!+\!1)$ with cost $(0,1)$
        \item Move: $(v,t)\!\rightarrow\!(u,t\!+\!1)$ for $u\!\in\!N(v)$ with cost $(1,1)$
    \end{itemize}
    Here, $N(v)$ denotes the neighborhood of $v$. A successor is only valid, if it can be reached without collision according to the reservation table and if the traversed edge is not defective.
    \item \emph{Cost Function}: Each node $n$ maintains the accumulated cost $g(n)=(moves, time)$ where $g(n)$ represents the total cost incurred from the start node up to node $n$.
    Costs are compared lexicographically, prioritizing fewer moves over shorter time. 
    This allows the search to prefer waiting over taking unnecessary detours.
    \item \emph{Heuristic}: By labeling each node in the graph with a coordinate, we choose the Chebyshev distance~\cite{Cantrell.2000} as an heuristic for assessing the distance from a node $n$ to the target node: $h(n)=\max(\lvert x_n-x_\text{target}\rvert, \lvert y_n-y_\text{target}\rvert)$.
    \item \emph{Evaluation Function}: Assembling these components, the evaluation key of the $A^*$ search is:
    $$f(n)=(g_\text{moves}(n)+h(n), g_\text{time}(n)+h(n))$$
\end{itemize}

All interactions for which no collision-free paths can be found, that satisfy the rules from \autoref{sec:constraints}, are deferred to a new layer, which is inserted immediately after the current layer. 
This allows the algorithms to retry routing these interactions in a reduced interaction set. 
We refer to this newly created layer as the \emph{spillover layer}.
If routing fails for all interactions in a spillover layer, the algorithm repeatedly explores alternative, possibly non-optimal meeting and target node assignments until either a feasible path is found or a predefined retry threshold is exceeded.

\subsection{Qubit Relocation}
For rotation algorithms, relocating blocking idle qubits is straightforward, since they are simply rotated along the diamonds according to a predefined pattern.
For path algorithms, however, relocation is more complex because idle qubits must be moved in a way that avoids collisions with other qubits. 
This issue can be addressed by reusing the path-computation procedure introduced in the previous section, with suitable constraints to account for the occupied nodes. 
For each idle qubit, the nearest available stabilizer node is chosen such that it is neither the start nor the target node of any active qubit and has not already been assigned to another idle qubit. 
Collision-free relocation paths are then computed using the same algorithm as described in \autoref{sec:path-comp}, while excluding all nodes currently occupied by active qubits. 
In the special case where a blocking idle qubit is itself obstructed by other non-blocking idle qubits, an additional recursive relocation procedure is required to clear the way for the active qubits.
This process is illustrated in the following example.
\begin{figure}[tb]
    \centering
    \includegraphics[width=0.95\linewidth]{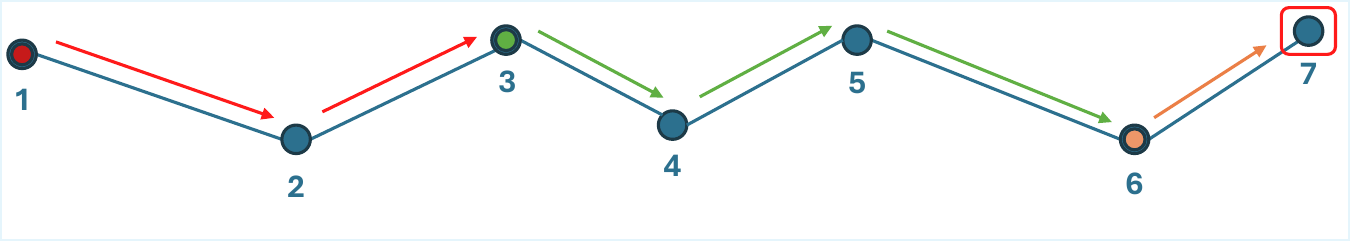}
    \caption{The recursive relocation process.}
    \label{fig:idle-evacuation}
\end{figure}
\begin{description}[topsep=6pt, leftmargin=0pt]
    \item[Example 14.] \itshape Consider the scenario illustrated in \autoref{fig:idle-evacuation}. Here, nodes 1, 2, 4, and 5 are part of the paths computed for active qubits. Along the relocation path of the red blocking idle qubit to its destination (node 7), additional non-blocking idle qubits (green and orange) obstruct its routing. To resolve this situation, the idle qubits are relocated recursively, indicated by the colored arrows, until node 1 becomes available.
\end{description}
If the path of some active qubits cannot be freed, e.g., due to a lack of available stabilizer nodes, the blocked qubits are placed into a spillover layer and executed separately.

For further technical details, we refer to the open-source implementation of our framework, which is publicly available as part of the \emph{Munich Quantum Toolkit} (MQT)~\cite{Wille.2024} at \mbox{\url{https://github.com/munich-quantum-toolkit/spin-qubit-routing}}.

%% file: sections/50_evaluation.tex
\vspace{-3mm}
\section{Evaluation}
\label{sec:evaluation}
With the compilation framework for the snakes on a plane model in place, we now evaluate its practical performance. 
Although the previous sections described the framework’s structure and algorithmic components, its behavior under different architectural constraints and problem sizes has not yet been quantified. 
The two routing paradigms introduced in \autoref{sec:path-and-rotation}, combined with different strategies for defect handling and initial mapping, lead to markedly different \mbox{trade-offs} in \emph{execution time} and \emph{routing overhead}. 
We therefore assess the framework using two complementary metrics: required timesteps, which capture execution time, and qubit movements, which reflect routing overhead. 
Based on these metrics, the evaluation is guided by the following research questions:

\begin{description}
    \item[RQ1 (Qubit Density):] How do path and rotation algorithms compare in terms of execution time and routing overhead as qubit density increases?
    \item[RQ2 (Adaptive Defect Handling):] To what extent do dynamic adaption strategies improve robustness and execution time, and what drawbacks do they introduce in defective architectures?
    \item[RQ3 (Initial Qubit Mapping):] How strongly does the choice of initial qubit mapping influence the routing performance of the proposed algorithms?
    \item[RQ4 (Scalability):] How do the computational complexities of path and rotation algorithms scale with increasing input size, measured by the number of qubits?
\end{description}

\subsection{Experimental Setup}
To answer these research questions, we evaluated the proposed compilation framework on randomly generated quantum circuits composed exclusively of a given sequence of \mbox{two-qubit} interactions, with a fixed depth of 10 layers.
Unless stated otherwise, the number of qubits is chosen as 9, which turns out to be the densest possible occupancy of the grid where all algorithms remain reliable (see \autoref{sec:rq1}).
All experiments are conducted on an architecture grid with 24 stabilizer nodes as shown in \autoref{fig:path-scheme}. 
This grid is large enough to provide a non-trivial routing problem while still being small enough to support extensive repeated evaluation.
The default hardware parameters are set to a defect probability of $p_\text{defect}=1\%$ and a recovery probability of $p_\text{recovery}=25\%$. 
The values are chosen to represent a low but non-negligible fault rate characteristic of realistic near-term quantum hardware. 
The moderate recovery rate allows defects to persist long enough to affect execution, while still enabling gradual mitigation over time.
To reduce the impact of measurement variability, results are averaged over 100 randomly generated sequences for each experiment. 
In addition, every computed schedule is verified to be collision-free and consistent with the routing constraints introduced in \autoref{sec:graph-and-problem}.
To prevent unbounded waiting in defective architectures, all algorithms use a retry threshold of 50 consecutive routing attempts. 
Runs that exceed this threshold are counted as failures.
We conduct a series of four experiments, each designed to address one of the research questions above.

\subsection{RQ 1: Impact of Qubit Density}
\label{sec:rq1}
\begin{figure}[tb]
    \centering
    \includegraphics[width=\linewidth]{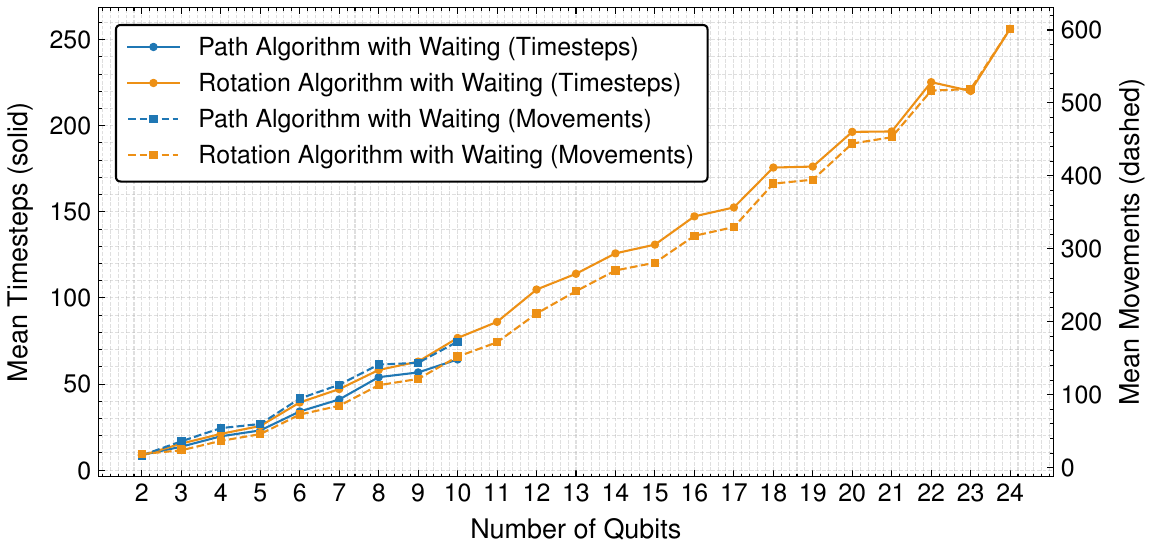}
    \caption{Comparison of path and rotation schemes as qubit density increases.}
    \label{fig:eval1}
\end{figure}

To analyze how path and rotation algorithms perform under increasing congestion, we vary the number of qubits while keeping all other parameters to their default values. 
Since defective edges are not the limiting factor in this experiment, we compare the path algorithm with waiting and the rotation algorithm with waiting, as no specialized defect-handling mechanisms are required. 
The results of this experiment are shown in \autoref{fig:eval1}.
Blue curves represent the path algorithm with waiting, while orange curves correspond to the rotation algorithm with waiting. 
Timesteps and movements are shown in the same plot, with timesteps being displayed on the left $y$-axis (solid lines) and movements on the right $y$-axis (dashed lines). 
The two main observations are:
\begin{itemize}
    \item The path algorithm requires more qubit movements, whereas the rotation algorithm incurs more timesteps.
    \item The rotation algorithm is more scalable than the \mbox{path-based} approach, maintaining functionality even under full grid occupancy, whereas the path algorithm yields results only for systems with up to 10 qubits within the retry limit.
\end{itemize}
These differences can be explained by the underlying mechanisms of the two approaches. 
The path algorithm exhibits a consistently higher movement count because it explicitly relocates idle qubits to clear routing paths. 
While this enables parallel execution once paths are established, resulting in relatively low timesteps, its dependence on available free stabilizer nodes makes it increasingly susceptible to congestion, ultimately leading to routing failures as the number of qubits grows.
In contrast, the rotation algorithm avoids explicit relocation by leveraging cyclic rotations to dynamically free nodes. 
This allows for more direct routing and eliminates the costly detours typically associated with idle qubit relocation in path algorithms. 
As a result, the approach remains functional even under full occupancy, making it inherently more scalable.
However, this advantage comes at the cost of frequent snake surgery teleportations at each movement step, which might introduce a technical overhead in practical implementations due to the required measurements.
Apart from that, the reliance on cyclic rotations creates sequential dependencies (see \autoref{sec:constraints}), increasing the total number of timesteps and limiting parallelism.

\subsection{RQ 2: Impact of Adaptive Defect Handling}
\begin{figure}[tb]
    \centering
    \includegraphics[width=\linewidth]{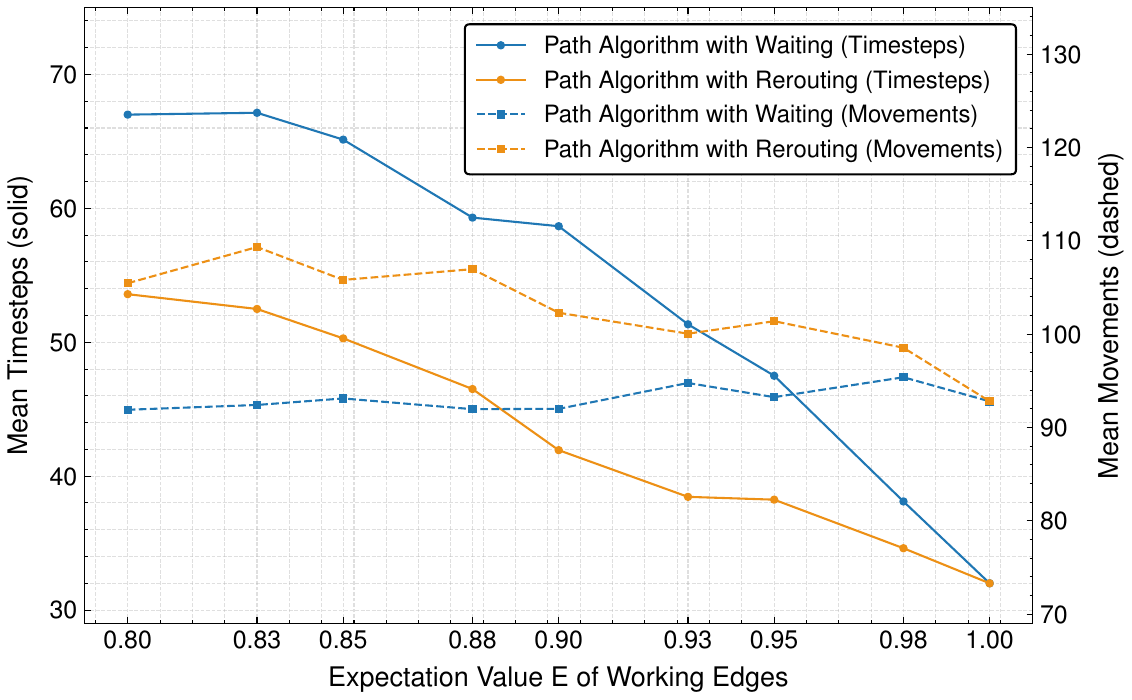}
    \caption{Effects of dynamic adaption on path algorithms.\vspace{1em}}
    \label{fig:eval2-3}
\end{figure}
\begin{figure}[tb]
    \centering
    \includegraphics[width=\linewidth]{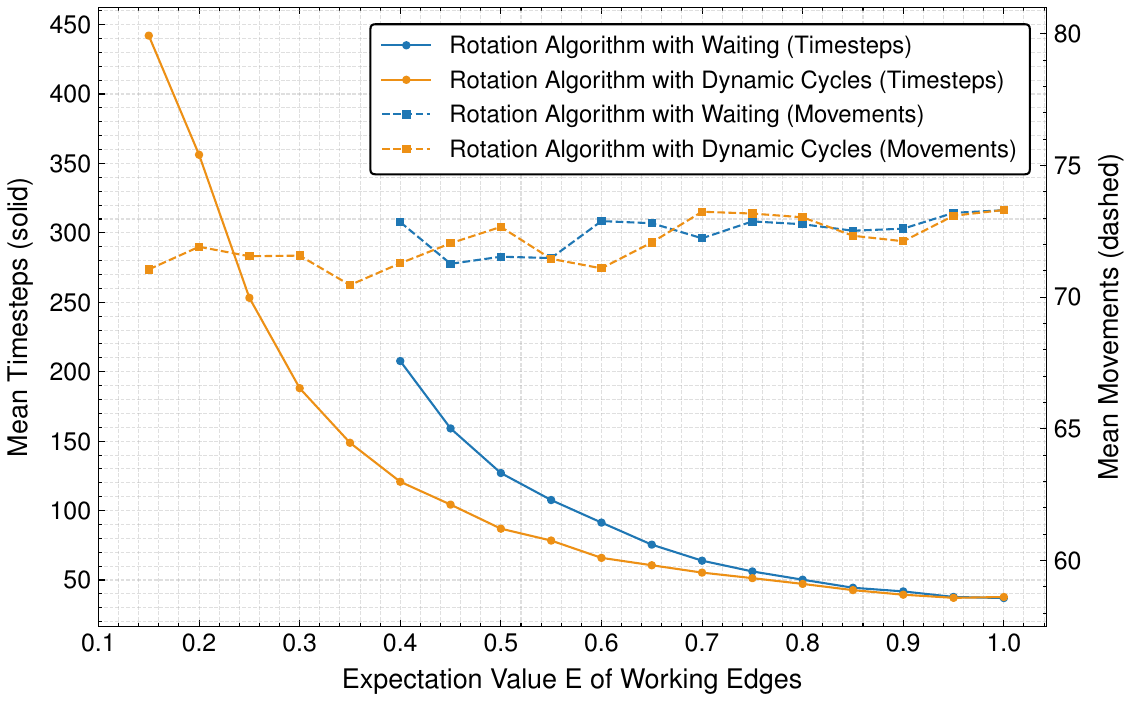}
    \caption{Effects of dynamic adaption on rotation algorithms.}
    \label{fig:eval2-2}
\end{figure}
To assess the benefits and limitations of dynamic adaption strategies, we vary the amount of defects occurring on the device and evaluate the performance for both waiting-based and dynamically adaptive variants across both algorithm classes.
Instead of considering $p_\text{defect}$ and $p_\text{recovery}$ independently, we express their combined effect through a single parameter $E$, denoting the expected fraction of functional edges.
The results for path algorithms are shown in \autoref{fig:eval2-3} and the results for rotation algorithms are illustrated in \autoref{fig:eval2-2}.
From these results, the following key messages can be derived:
\begin{itemize}
    \item Dynamic adaption reduces the number of required timesteps but increases the number of qubit movements for path algorithms.
    \item For rotation algorithms, the number of timesteps is also reduced by dynamic adaption but the differences in routing overhead are minimal.
    \item Dynamic adaptation further improves the robustness of rotation algorithms, which are already more robust than path algorithms.
\end{itemize}
These observations can be attributed to how each algorithm employs adaptive strategies in the presence of defects. In general, dynamic adaptation enhances robustness by actively exploiting alternative routing options, rather than passively waiting for defects to resolve.
For rotation algorithms, this adaptability is realized through dynamic cycle extension. 
By forming larger cycles, qubits can bypass defective edges without requiring a large amount of additional movements in most cases. 
As a result, valid routing schedules can still be computed even at higher defect rates, which explains the improved robustness. 
Furthermore, reducing idle waiting contributes to fewer overall timesteps.
In contrast, path algorithms rely on rerouting, which provides more limited benefits in terms of robustness, as the number of feasible alternative paths is typically smaller than the number of possible cycle adaptations. 
Additionally, when rerouting is feasible, it often introduces detours from optimal paths, leading to an increase in qubit movements. 
Nevertheless, similar to rotation-based approaches, actively exploring alternative routes helps to reduce idle waiting and thus lowers the total execution time.

\subsection{RQ 3: Impact of Initial Qubit Mapping}
\autoref{tab:initial_mapping_results} summarizes the average number of timesteps and movements for path and rotation algorithms with waiting under default conditions using the three mapping strategies included in our framework. Path algorithms were evaluated on 9 qubits, while rotation algorithms used a fully occupied grid.
\begin{table}[tb]
    \vspace{0.8em}
    \centering
    \small
    \resizebox{\linewidth}{!}{%
    \begin{tabular}{l@{\hspace{1.5em}} l@{\hspace{1.5em}} l@{\hspace{1.5em}} l}
        \toprule
        \textbf{Algorithm} & \textbf{Mapping Strategy}
        & \textbf{$\varnothing$ Timesteps} & \textbf{$\varnothing$ Movements} \\
        \midrule
        \multirow{3}{*}{Path Algorithm}
            & Random Mapping                & 64.28 (baseline) & 172.75 (baseline) \\
            & Reverse Traversal Mapping     & 62.35$\; \textcolor{ForestGreen}{(\downarrow 3.00\%)}$  & 161.35$\; \textcolor{ForestGreen}{(\downarrow 6.60\%)}$ \\
            & Interaction-based Mapping     & 63.76$\; \textcolor{ForestGreen}{(\downarrow 0.81\%)}$ & 166.53$\; \textcolor{ForestGreen}{(\downarrow 3.60\%)}$ \\
        \midrule
        \multirow{3}{*}{Rotation Algorithm}
            & Random Mapping                & 256.28 (baseline) & 601.58 (baseline) \\
            & Reverse Traversal Mapping     & 246.31$\; \textcolor{ForestGreen}{(\downarrow 3.89\%)}$ & 576.01$\; \textcolor{ForestGreen}{(\downarrow 4.25\%)}$ \\
            & Interaction-based Mapping     & 247.08$\; \textcolor{ForestGreen}{(\downarrow 3.59\%)}$ & 568.13$\; \textcolor{ForestGreen}{(\downarrow 5.56\%)}$ \\
        \bottomrule
    \end{tabular}%
    }
    \caption{Evaluation of initial-mapping strategies.}
    \label{tab:initial_mapping_results}
\end{table}

Reverse traversal mapping is especially effective because it propagates the impact of earlier gates in reverse execution order, resulting in initial qubit placements that better anticipate future routing requirements. 
In contrast, \mbox{interaction-based} mapping places frequently interacting qubits close together, thereby reducing routing distances and minimizing unnecessary shuttling.
Together, these approaches highlight how incorporating the interaction sequence can significantly improve the benefits of initial mapping.

\subsection{RQ 4: Scalability Analysis}
\begin{figure}[tb]
    \centering
    \includegraphics[width=\linewidth]{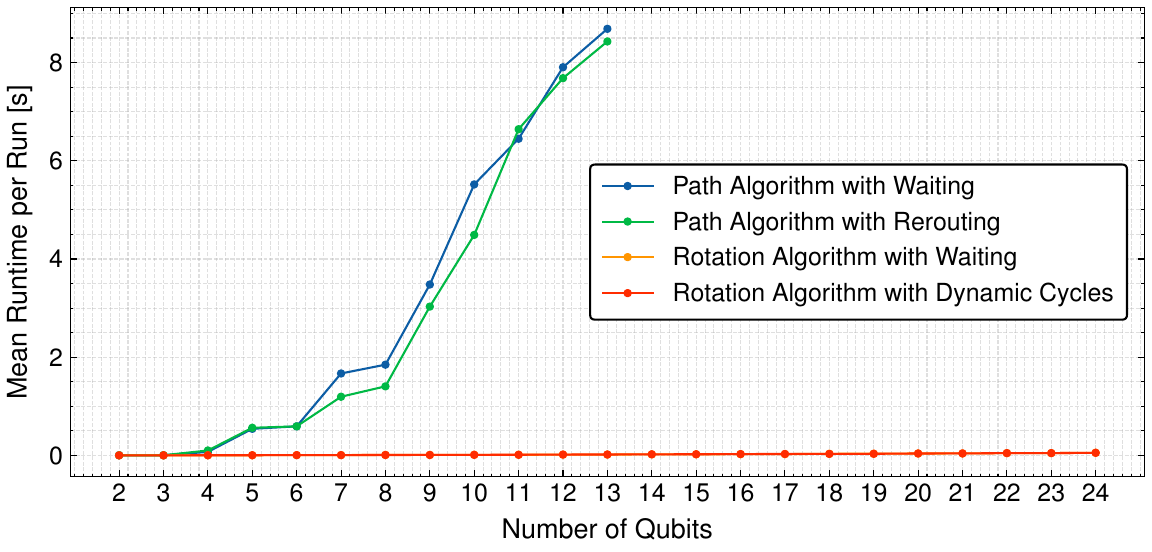}
    \caption{Impact of qubit count on CPU runtime.}
    \label{fig:eval-runtime}
\end{figure}

To analyze the impact of qubit density on runtime, we systematically vary the number of qubits and measure the average runtime per configuration across all runs for each of the four algorithms.
The results in \autoref{fig:eval-runtime} show the following:
\begin{itemize}
    \item The runtime of path algorithms increases with qubit density, while the runtime of rotation algorithms remains nearly constant at a very low level.
    \item The path algorithm with waiting exhibits a slightly higher runtime than the rerouting variant.
\end{itemize}
The higher runtime complexity of path algorithms stems from the need to explicitly compute collision-free paths for both active and idle qubits. 
In contrast, the rotation algorithms only compute paths for active qubits and subsequently handle idle qubits by applying rotations according to a predefined pattern.
Furthermore, the slightly increased runtime of the path algorithm with waiting is due to its greater number of waiting cycles compared to the path algorithm with rerouting. 
Since it does not consider alternative paths, it requires additional algorithm iterations.

\subsection{Implications and Practical Trade-Offs}
This evaluation shows that path algorithms and their implementations might not scale to denser architectures, especially at high qubit densities.
For such architectures, one should consider using the rotation-based approach or a modified path algorithm that keeps the path-computation complexity within certain bounds, at the cost of optimality.
The development of such large-scale algorithms will be future work, based on the insights from the routing methods studied in this work.

%% file: sections/60_conclusion.tex
\section{Conclusion}
\label{sec:conclusion}
In this work, we presented a compilation framework for spin qubit quantum computers based on the snakes on a plane architecture, modeling qubit shuttling as a constrained routing problem on a graph. Within this framework, we developed two routing strategies: path algorithms, which rely on explicit collision-free paths, and rotation algorithms, which use cyclic rotations to resolve congestion.
Our evaluation shows complementary trade-offs between both approaches. 
Path algorithms achieve lower execution times in well-connected, \mbox{low-defect} settings but are sensitive to congestion and defects.
In contrast, rotation algorithms remain robust under high qubit density and limited connectivity, at the cost of longer execution times, while requiring fewer qubit movements.
We further demonstrated that adaptive defect-handling improves performance by reducing execution time, and that initial qubit mapping also has a large influence on routing efficiency.
Together, this work constitutes one of the first compilation methods for error-corrected spin qubit architectures.
Future research will focus on developing path-based approaches for dense \mbox{large-scale} architectures that require a reduced number of snake teleportations compared to the rotation algorithms.

\section*{Acknowledgements \& AI Usage Disclosure}
This work received funding from the European Research Council (ERC) under the European Union’s Horizon 2020 research and innovation program (grant agreement No. 101001318) and was part of the Munich Quantum Valley, which is supported by the Bavarian state government with funds from the Hightech Agenda Bayern Plus.
Furthermore, it was funded by the Deutsche Forschungsgemeinschaft (DFG, German Research Foundation, No. 563402549).

During the preparation of this manuscript, AI-based language models were used to improve readability, spelling, grammar, and overall clarity. 
All outputs generated by these tools were carefully reviewed and edited manually as needed. 
The authors take full responsibility for the final content.